\documentclass[conference]{IEEEtran}
\IEEEoverridecommandlockouts
\usepackage{cite}
\usepackage{amsmath,amssymb,amsfonts}
\usepackage{algorithmic}
\usepackage{graphicx}
\usepackage{textcomp}
\usepackage{xcolor}
\usepackage{enumitem}
\usepackage{bm} 
\usepackage{algorithmic}
\usepackage{algorithm}
\usepackage{mathrsfs}

\usepackage{makecell}  
\usepackage{amsmath}   

\usepackage[top=2.5cm, bottom=2.5cm, left=2.5cm, right=2.5cm]{geometry}

\usepackage{stfloats}
\newcommand{\figureref}[1]{Fig.~\ref{#1}}
\def\BibTeX{{\rm B\kern-.05em{\sc i\kern-.025em b}\kern-.08em
    T\kern-.1667em\lower.7ex\hbox{E}\kern-.125emX}}
\begin{document}

\title{Efficient Precoding in XL-MIMO-AFDM System\\
\thanks{This paper is supported in part by National Natural Science Foundation of China Program(62271316, 62101322), National Key R\&D Project of China (2019YFB1802703), Shanghai Key Laboratory of Digital Media Processing (STCSM 18DZ2270700) and the Fundamental Research Funds for the Central Universities.

The authors: Jun Zhu, Yin Xu, Dazhi He, Haoyang Li, YunFeng Guan and Wenjun Zhang are from Shanghai Jiao Tong University. Dazhi He is from Pengcheng Laboratory. The corresponding author is Dazhi He (e-mail: hedazhi@sjtu.edu.cn).
}
}

\author{
    \IEEEauthorblockN{Jun Zhu\IEEEauthorrefmark{1}, Yin Xu\IEEEauthorrefmark{1}, Dazhi He\IEEEauthorrefmark{1}\IEEEauthorrefmark{3}, Haoyang Li\IEEEauthorrefmark{1}, Yunfeng Guan\IEEEauthorrefmark{2}, Wenjun Zhang\IEEEauthorrefmark{1}, \textit{Fellow, IEEE},\\  Tianyao Ma \IEEEauthorrefmark{1}, Haozhi Yuan \IEEEauthorrefmark{1} }
    \IEEEauthorblockA{\IEEEauthorrefmark{1} Cooperative Medianet Innovation Center (CMIC), Shanghai Jiao Tong University\\\IEEEauthorrefmark{3}Pengcheng Laboratory \\Shanghai 200240, China \\ Email: \{zhujun\_22,  xuyin, hedazhi,  lihaoyang,  zhangwenjun,  mty0710, yuanhaozhi\}@sjtu.edu.cn\\\IEEEauthorrefmark{2} Institute of Wireless Communication Technology, College of Electronic Information and Electrical Engineering \\ Email: yfguan69@sjtu.edu.cn}
}

\maketitle

\begin{abstract}
This paper explores the potential of afﬁne frequency division multiplexing (AFDM) to mitigate the multiuser interference (MUI) problem by employing time-domain precoding in extremely-large-scale multiple-input multiple-output (XL-MIMO) systems. In XL-MIMO systems, user mobility significantly improves network capacity and transmission quality. Meanwhile, the robustness of AFDM to Doppler shift is enhanced in user mobility scenarios, which further improves the system performance. However, the multicarrier nature of AFDM has attracted much attention, and it leads to a significant increase in precoding complexity. However, the serious problem is that the multicarrier use of AFDM leads to a sharp increase in precoding complexity. Therefore, we employ efficient precoding randomized Kaczmarz (rKA) to reduce the complexity overhead. Through simulation analysis, we compare the performance of XL-MIMO-AFDM and XL-MIMO orthogonal frequency division multiplexing (XL-MIMO-OFDM) in mobile scenarios, and the results show that our proposed AFDM-based XL-MIMO precoding design can be more efficient.

\end{abstract}


\section{Introduction}
\IEEEPARstart{E}{xtremely}-large-scale multiple-input multiple-output (XL-MIMO) communication systems have emerged as a promising technology for 6G, offering significant potential for future communication systems\cite{b1}, \cite{b2}. Unlike conventional massive MIMO systems, which typically assume a planar wavefront, XL-MIMO is able to increase the spacing between antennas to significantly increase the likelihood of near-field propagation scenarios. This enhancement allows for the frequent applicability of the line-of-sight (LoS) assumption in mobile environments, as scattering and reflections are less prominent. However, mobility introduces challenges such as timing offsets, phase noise, and Doppler frequency offset (DFO), all of which can lead to severe inter-carrier interference (ICI) and degrade overall system performance\cite{b3}, \cite{b4}. In this context, the XL-MIMO architecture presents a promising solution, effectively addressing the issues posed by mobile environments and paving the way for more robust and efficient communication systems in the future.

It is important to note that the current design of XL-MIMO systems for mobile scenarios primarily relies on XL-MIMO-OFDM \cite{b5}, \cite{b6}. However, this approach is not always optimal. An alternative and promising solution is affine frequency division multiplexing (AFDM), which effectively isolates propagation paths with distinct delays or Doppler shifts under LTV channels within the one-dimensional Discrete Affine Fourier Transform (DAFT) domain. By strategically adjusting the linear frequency modulation (FM) slope in AFDM to align with the channel’s Doppler profile, this technique offers a more targeted and efficient method for addressing Doppler-induced distortions \cite{b7, b8, b9, b10}.

In XL-MIMO systems, minimizing the impact of multi-user interference (MUI) on system performance has been a primary concern. Given the nature of ultra-large-scale transmit antenna arrays, most existing research has focused on low-complexity precoding techniques. The randomized Kaczmarz (rKA) algorithm, as proposed in [11], significantly reduces the computational overhead of the system. Additionally, the substitution-free sampling rKA (SwoR-rKA) algorithm introduced in [12] enhances system performance while maintaining low complexity. However, research on MUI mitigation within the context of MIMO-AFDM remains limited, primarily due to the increased complexity of precoding design in AFDM. Moreover, most studies still prioritize low-complexity solutions. Notably, [13][14] have demonstrated that low-complexity zero-forcing (ZF) precoding is particularly advantageous for mitigating the multiuser interference (MUI) in MIMO-AFDM systems.

In recent years, XL-MIMO and AFDM have attracted a lot of attention in the field of communications.The near-field nature of XL-MIMO makes it possible to design for high-speed mobile scenarios. Meanwhile, the application of AFDM can be more robust to DFO. Massive transmit antenna arrays can provide good gain, but there are also complexity and multi-user interference issues that need to be addressed. In XL-MIMO systems, the use of low-complexity precoding is important and mitigates inter-user interference to some extent. According to the characteristics of XL-MIMO systems, AFDM can have better robustness to DFO. However, the subcarrier application of AFDM makes the precoding design complexity a catastrophic problem.

In this paper, in contrast to the studies presented in \cite{b5}, \cite{b13}, and \cite{b14}, We propose a more efficient method to make Doppler shift more robust, reduce MUI and reduce system complexity. We have established an accurate MUI model for the XL-MIMO-AFDM system and designed an optimization function aimed at maximizing system performance while minimizing computational complexity. To address the Doppler shift issue in the XL-MIMO mobile scenario, we adopted the XL-MIMO-AFDM system model and combined it with the rKA precoding technique to effectively suppress MUI while ensuring low complexity. This approach significantly improves communication performance in dense user access environments in XL-MIMO mobile scenarios. Through simulation experiments, we fully verified the feasibility and effectiveness of the proposed method, demonstrating its potential for practical applications and providing valuable insights for the design of efficient communication systems.

 The rest of the paper is composed as follows. In Section \text{II}, the XL-MIMO-AFDM model is introduced. In Section \text{III}, efficient precoding is introduced and the complexity overhead of the proposed precoding is analyzed. In Section \text{IV}, simulation results show that AFDM is  robust to DFO and the proposed algorithm has lower complexity than XL-MIMO-OFDM. In Section \text{V}, the conclusions of this paper are summarized.

Notation: Bold uppercase letters denote matrices and bold lowercase letters denote vectors. For a matrix $\textbf A$, $\textbf A^T$, $\textbf A^H$, $\textbf A^{-1}$, and Tr($\textbf A$) denote its transpose, conjugate transpose, inverse, and trace, respectively. blkdiag($\textbf H_1, ..., \textbf H_K$) denotes a block diagonal matrix with $\textbf H_1, ..., \textbf H_K$ being its diagonal blocks. The space of M × N complex matrices
is expressed as $\mathbb{C}^{M\times N}$.

\section{SYSTEM MODEL}

\begin{figure}[t]
\centering
\includegraphics[width=0.5\textwidth]{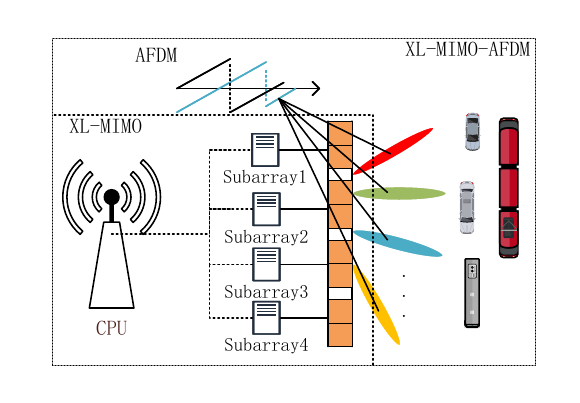}
\caption{ XL-MIMO-AFDM System.}
\label{fig:1}
\end{figure}

In this section, we introduce the system model of XL-MIMO-AFDM, including XL-MIMO and AFDM. Based on \figureref{fig:1}, we know that the channel model employs a dynamic scenario.
\subsection{AFDM Channel Model}
\begin{figure}[t]
\centering
\includegraphics[width=0.5\textwidth]{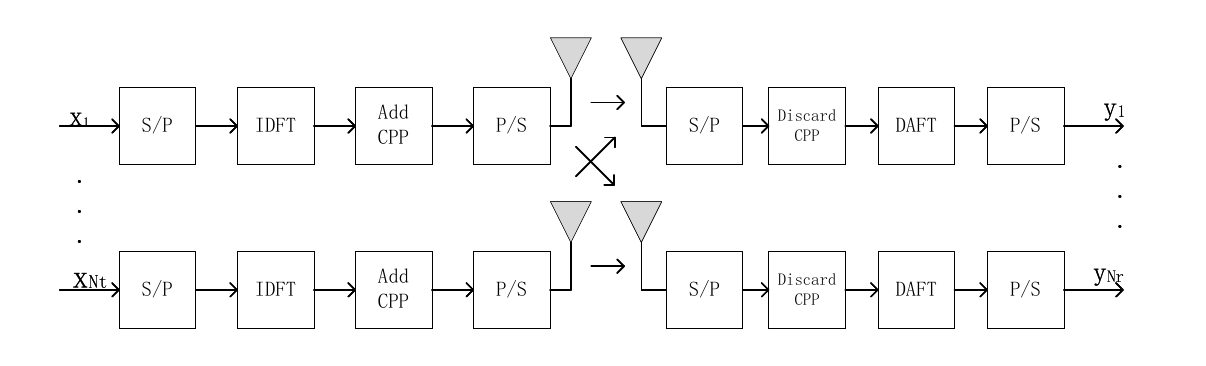}
\caption{MIMO-AFDM modulation/demodulation block diagrams.}
\label{fig:2}
\end{figure}
The AFDM modulation scheme is originally based on DAFT, such that the AFDM signal vector $\textbf s \in {\mathbb{C}^{{N}\times1}}$ ($N$ being the number of chirp carriers) can be expressed as
\begin{equation}
\begin{aligned}
\textbf s = \mathbf{\Lambda }_{c_1}^H \textbf F^H \mathbf{\Lambda }_{c_2}^H\textbf x=\textbf A\textbf x,
         \end{aligned}
\label{eq:1}
\end{equation}
\begin{figure*}[!hb]
	\hrulefill
	\vspace*{5pt}
\begin{align}
     y(m) = \frac{1}{N}\sum_{i=1}^{P}\sum_{m'=1}^{N-1} h_i e^{j\frac{2\pi}{N}(Nc_1l_i^2-m'li+c_2(m'^2-m^2))\frac{e^{-j2\pi(m+ind_i-m'+\beta_i)-1}}{e^{-j\frac{2\pi}{N}(m+ind_i-m'+\beta_i)-1}}} x(m')+N(m).  
\tag{2}  
\label{eq:2}
\end{align}
\begin{align}
   SINR_{jk}^s = \frac{|{\textbf h_{jk}^j}^H\textbf F_{jk}|^2}{{\sum_{i\neq k}^{K_s}|{\textbf h_{jk}^{j}}^H\textbf F_{ji}|^2 }+{\sum_{s\neq j}^{S}\sum_{i=1}^{K_s}|{\textbf h_{jk}^{s}}^H\textbf F_{jk}^s|^2}+\sigma_{jk}^2} 
\tag{23}  
\label{eq:23}.
\end{align}
\end{figure*}
where $\mathbf{\Lambda_c}\triangleq diag(e^{-j2\pi cn^2},n = 0, 1, ..., N-1)$, $c_1$ and $c_2$ are two AFDM parameters, $\textbf F$ is the DFT matrix with $e^{-j2\pi mn/N}/\sqrt{N}$, $\textbf A = \mathbf{\Lambda_{c_1}^H} \textbf F^H \mathbf{\Lambda_{c_2}^H} {\mathbb{C}^{{N}\times N}} $represents the DAFT matrix, $\textbf x \in \mathbb{A}^{N \times 1}$ denotes a vector of $N$ quadrature amplitude modulation (QAM) symbols that reside on the DAFT domain, $\mathbb{A}$ represents the modulation alphabet. We have the input-output relationship of SISO-AFDM system in the DAFT domain as shown in \eqref{eq:2} \cite{b9}. We can know that

\begin{equation}
\begin{aligned}
 c(l_i,m,m') = e^{j\frac{2\pi}{N}(Nc_1l_i^2-m'l_i+c_2(m'^2-m^2))}
         \end{aligned}
         \tag{3}  
\label{eq:3},
\end{equation}

\begin{equation}
\begin{aligned}
&F(l_i,v_i,m,m') = \frac{e^{-j2\pi(m+ind_i-m'+\beta_i)-1}}{e^{-j\frac{2\pi}{N}(m+ind_i-m'+\beta_i)-1}}\\
&ind_i = (\alpha_i+2Nc_1l_i)_N
         \end{aligned}
         \tag{4}  
\label{eq:4},
\end{equation}
where non-negative integer $l_i \in[0, l_{max}]$ is the associated delay of the $i-$th path with the $T_s$ normalization, $m $ denote the time and DAFT domains indices, $m\in{[0,N-1]}, r\in{[0,N_r]}$, $v_i = \alpha_i + \beta_i$ represents the associated Doppler shift normalized with subcarrier spacing and has a finite support bounded by $[-v_{max}, v_{max}]$, $ind_i$ denotes the index indicator of the i-th path, $ \alpha_i\in [-\alpha_{max}, \alpha_{max}]$ and $\beta_i \in (-\frac{1}{2} , \frac{1}{2} ]$ are the integer and fractional parts of $v_i$ respectively, $v_{max}$ denotes the maximum Doppler and $\alpha_{max}$ denotes its integer component. In this paper, we assume that $l_{max}$ and $ v_{max}$ are known in advance.  

According to \eqref{eq:2}\cite{b9}, we can know the channel model of MIMO-AFDM. Where $N_t$ and $N_r$ represent the number of transmit antennas (TA) and receive antennas (RA), respectively.
\begin{equation}
\begin{aligned}
 y_r(m) =& \sum_{t=1}^{N_t}\sum_{i=1}^{P}\sum_{m'=1}^{N-1} \frac{1}{N}h_i^{(r,t)}c(l_i,m,m') \\
         &\quad \times F(l_i,v_i,m,m')x_t(m)+W_r(m),
         \end{aligned}
         \tag{5}  
\label{eq:5}
\end{equation}
where  $P$ is the number of paths, $t\in{[0,N_t]}$ denote the index of the RA and TA respectively, $h_i^{(r,t)}$ is the channel gain of the $i-$th path between the $r-$th RA and the $t-$th TA, $W_r \in \mathcal{CN}(0, \sigma_k^2)$ represents the noise in DAFT domain at the $r-$th RA.

\eqref{eq:5} can be written as
\begin{equation}
\begin{aligned}
\textbf y_1 =  &\textbf H_{1,1}\textbf x_{1,1}+\textbf H_{1,2}\textbf x_{1,2}+\dots+\textbf H_{1,N_t}\textbf x_{1,N_t}+\textbf w_1\\
 &\vdots\\
\textbf y_{N_r} = & \textbf H_{N_r,1}\textbf x_{N_r,1}+\textbf H_{N_r,2}\textbf x_{N_r,2}+\dots+\textbf H_{N_r,N_t}\textbf x_{N_r,N_t}\\&+\textbf w_{N_r}
         \end{aligned}
         \tag{6}  
\label{eq:6},
\end{equation}

\begin{equation}
\begin{aligned}
\textbf H_{r,t} = \sum_{i=1}^Ph_i^{(r,t)}\textbf H_i
         \end{aligned}
         \tag{7}  
\label{eq:7},
\end{equation}

\begin{equation}
\begin{aligned}
\textbf H_i = \frac{1}{N}c(l_i,m,m')F(l_i,v_i,m,m')
         \end{aligned}
         \tag{8}  
\label{eq:8}.
\end{equation}

We define the effective MIMO channel matrix for the above MIMO-AFDM system as
\begin{equation}
\begin{aligned}
   \textbf H_{MIMO} = 
\begin{bmatrix}
 \textbf H_{1,1}&\dots&& \textbf H_{1,N_t}  \\
\vdots&\ddots \\
 \textbf H_{N_r,1}&\dots&& \textbf H_{Nr,N_t} \\
\end{bmatrix}.
\end{aligned}
\tag{9}  
\label{eq:9}
\end{equation}

That \eqref{eq:6} can be rewritten as
\begin{equation}
\begin{aligned}
\textbf y_{MIMO} = \textbf H_{MIMO}\textbf x_{MIMO}+\textbf w_{MIMO}
         \end{aligned},
         \tag{10}  
\label{eq:10}
\end{equation}
where $\textbf y_{MIMO}=[\textbf y_1^T, \textbf y_2^T,\dots, \textbf y_{N_r}^T]^T\in{\mathbb{C}^{NN_r\times 1}}$, $\textbf H_{MIMO}\in{\mathbb{C}^{NN_r\times NN_t}}$, $\textbf x_{MIMO}=[\textbf x_1^T, \textbf x_2^T,\dots, \textbf x_{N_t}^T]^T\in{\mathbb{C}^{NN_t\times 1}}$, $\textbf w_{MIMO}=[\textbf w_1^T, \textbf w_2^T,\dots, \textbf w_{N_r}^T]^T\in{\mathbb{C}^{NN_r\times 1}}$.
\subsection{XL-MIMO-AFDM Channel Model}
As shown in Figure 1, we investigate the downlink of an XL-MIMO-AFDM system where the base station (BS) is equipped with $N_t$ antennas. These antennas are divided into S subarrays, with each subarray having $N_{ts} = \frac{N_t}{S}$ antennas. Notably, each subarray is equipped with its own Local Processing Units (LPUs) for signal processing, and all these LPUs are connected to a Central Processing Unit (CPU). Additionally, K users are evenly distributed across the S subarrays, with each subarray serving $K_s = \frac{K}{S}$  users. Each user is equipped with a receiving antenna, $N_r = \sum_{k=1}^KN_k$. Therefore, the received signal at the $k-$th user is
 
\begin{equation}
\begin{aligned}
\textbf y_{k} = \sum_{s=1}^S{\textbf h_{k}^{s}}^H\textbf x_s+\textbf w_{k},
         \end{aligned}
         \tag{11}  
\label{eq:11}
\end{equation}
where $\textbf y_k \in{\mathbb{C}^{NN_k\times NN_{ts}}}$ is the received
signal, $\textbf h_{k} = [(\textbf h_{k}^1)^T, (\textbf h_{k}^2)^T, \dots, (\textbf h_{k}^S)^T]^T\in{\mathbb{C}^{NN_k\times NNt}}$, $\textbf x = [\textbf x_1^T,\textbf x_2^T, \dots, \textbf x_S^T]^T$, $\textbf x_s \in{\mathbb{C}^{NN_{ts}\times 1}}$ is the transmit signal in the s-th subarray and $\textbf h_{k}^s $ represents the channel vector between the BS in the s-th subarray and the k-th UE in the j-th subarray, respectively, $Nr = \sum_{s=1}^SN_{rs}$.

\begin{equation}
\begin{aligned}
\textbf x_s = \sum_{i=1}^{K_s}\textbf f_{si}\textbf s_{si}=\textbf F_s\textbf s_s
         \end{aligned},
         \tag{12}  
\label{eq:12}
\end{equation}
where $\textbf F_s = [(\textbf f_{s1})^T, (\textbf f_{s2})^T, \dots, (\textbf f_{sK_s})^T]^T\in{\mathbb{C}^{NN_{ts}\times D_s}}$, $\textbf s_s = [(\textbf s_{s1})^T, (\textbf s_{s2})^T, \dots, (\textbf s_{sK_s})^T]^T\in{\mathbb{C}^{D_s\times 1}}$, $D = \sum_{s=1}^SD_s$, $D$ denotes the total data flow.
\begin{equation}
\begin{aligned}
\textbf y_{k}^j = &\underbrace{{\textbf h_{k}^{j}}^H\textbf F_{jk}\textbf s_{k}}_{\hspace{0cm} \text{Desired signal}}+\underbrace{\sum_{i\neq k}^{K_s}{\textbf h_{k}^{j}}^H\textbf F_{ji}\textbf s_{ji}}_{\hspace{0cm} \text{Intra-subarray interefence}}\\&+\underbrace{\sum_{s\neq j}^{S}\sum_{i=1}^{K_s}{\textbf h_{k}^{s}}^H\textbf F_{jk}^s\textbf s_{si}}_{\hspace{0cm} \text{Interefence-subarray interefence}}+\textbf w_{jk}
         \end{aligned}
         \tag{13}  
\label{eq:13},
\end{equation}
where $\textbf F_{s} = [\textbf F_{s1}^T, \textbf F_{s2}^T, ...,\textbf F_{sKs}^T]^T$ denote the channel matrix between the $j-$th subarry and user $k$, $\textbf F_{sKs}\in{\mathbb{C}^{NN_{ts}\times D_{sk}}}$, $D = \sum_{k=1}^{K_s}D_{sk} $ .

In this paper, we consider a channel model with the non-stationary property. The channel vector between UE $k$ in the j-th subarray and $N_{ts}$ antennas in the $s-$th subarray is denoted by
\begin{equation}
\begin{aligned}
\textbf h_{k}^s = \sqrt{N_{ts}}(\Theta_{jk}^s)^{\frac{1}{2}}\textbf z_{jk}^s\textbf H_{r,t}^{ks}
         \end{aligned},
         \tag{14}  
\label{eq:14}
\end{equation}
where $\textbf h_{k}^s\in{\mathbb{C}^{N_k\times NN_{ts}}}, \textbf z_{jk}^s\in \mathcal{CN}(0, \frac{1}{N_{ts}} \textbf I_{N_{ts}}), \textbf H_{r,t} = [(\textbf H_{r,t}^{1})^T, (\textbf H_{r,t}^{2})^T, \dots, (\textbf H_{r,t}^{K})^T]^T, \textbf H_{r,t}^k\in{\mathbb{C}^{N_k\times NN_{t}}}$ denotes the channel matrix of user $K$, $\textbf H_{r,t}^k = [(\textbf H_{r,t}^{k1})^T, (\textbf H_{r,t}^{k2})^T, \dots, (\textbf H_{r,t}^{kS})^T]^T$, $\textbf H_{r,t}^{kS}\in{\mathbb{C}^{N_k\times NN_{ts}}}$ denotes the channel matrix of user $K$ in $S-$th subarray.
\begin{equation}
\begin{aligned}
\Theta_{jk}^s = (\textbf D_{jk}^s)^{\frac{1}{2}}\textbf R_{jk}^s(\textbf D_{jk}^s)^{\frac{1}{2}}
         \end{aligned},
         \tag{15}  
\label{eq:15}
\end{equation}
where  $\textbf R_{jk} \in{\mathbb{C}^{N_{ts}\times N_{ts}}}$ is the spatial correlation matrix between UE $k$ in the $j-$th subarray and BS in the s-th subarray, $\textbf D_{jk} \in{\mathbb{C}^{N_{ts}\times N_{ts}}}$ is a diagonal matrix and has $\textbf D_{jk}^s$ non-zero diagonal elements between UE $k$ in the $j-th$ subarray and BS in the $s-$th subarray\cite{b12}.

\section{EFFICIENTLY PRECODING}
The rKA is an iterative algorithm that solves systems of linear equations and exhibits good performance, demonstrating relatively low complexity in ultra-large-scale transmit antenna array systems.
\subsection{Randomized Kaczmarz Signal Precoding}
Based on \eqref{eq:11} and \eqref{eq:12}, we can observe that traditional ZF and RZF have relatively high complexity in large-scale systems. rKA, used for solving linear equations, can maintain good performance while having lower complexity. The precoding matrix $G_j^{RZF}$ in the j-th subarray can be computed according to
\begin{equation}
\begin{aligned}
\textbf G_{j}^{RZF} =\beta \textbf H_{j}^j((\textbf H_{j}^j)^H\textbf H_{j}^j)^{-1}
         \end{aligned},
         \tag{16}  
\label{eq:16}
\end{equation}
According to \eqref{eq:16}, we can know that the received signal is
\begin{equation}
\begin{aligned}
\textbf x_{j}=\textbf G_j\textbf s_j =\beta \textbf H_{j}^j((\textbf H_{j}^j)^H\textbf H_{j}^j)^{-1}\textbf s_j
         \end{aligned},
         \tag{17}  
\label{eq:17}
\end{equation}

In \eqref{eq:18}, the computational complexity is relatively low when the number of transmitting antenna arrays is small. However, as the number of transmitting antenna arrays increases, the complexity escalates. To mitigate this issue, the rKA algorithm is utilized to effectively reduce the overall computational complexity.
\begin{equation}
\begin{aligned}
\textbf A_{j}=((\textbf H_{j}^j)^H\textbf H_{j}^j)^{-1}\textbf s_j
         \end{aligned},
         \tag{18}  
\label{eq:18}
\end{equation}
Exploiting the rKA algorithm, we can transform the former expression in \eqref{eq:18} into an optimization problem
\begin{equation}
\begin{aligned}
\textbf A_{j}=\underset{\textbf  x_{j}\in {\mathbb{C}^K}}{\arg\min}||\textbf H_j^j\textbf x_j||^2+\xi||\textbf x_j-\textbf s_j||^2
         \end{aligned},
         \tag{20}  
\label{eq:20}
\end{equation}
where $\textbf s_j^{\xi}=\frac{s_j}{\xi}$ is the transmitted signal combining the regularization factor in the j-th subarray.

By taking the derivative of $A_j$, we can obtain
\begin{equation}
\begin{aligned}
\textbf A_{j}=\underset{\textbf  x_{j}\in {\mathbb{C}^K}}{\arg\min}||\textbf B_j\textbf x_j-\textbf a_j||^2
         \end{aligned},
         \tag{20}  
\label{eq:20}
\end{equation}
where $\textbf B_j = [\textbf H_j^j]\in {\mathbb{C}^{{(N_{ts}+K)}\times K}}$, $\textbf a_j=[0,\textbf s_j]\in {\mathbb{C}^{{(N_{ts}+K)}}}$, The set of linear equations (SLE) $\textbf A_j\textbf x_j = \textbf a_j$ is over-determined (OD) and should be solved for the vector $\textbf x_j \in {\mathbb{C}^K    }$. This SLE is inconsistent unless $s_j = 0$, so the above SLE can be denoted
\begin{equation}
\begin{aligned}
\textbf A_{j}^H\textbf a_j=\textbf s_j
         \end{aligned},
         \tag{21}  
\label{eq:21}
\end{equation}
Based on the fundamental theory of rKA and \cite{b11}, \cite{b12}, we can obtain \eqref{eq:17}.
\begin{algorithm}[]
    \caption{ Efficiently Precoding in XL-MIMO-AFDM System}
	\label{algorithm1} 
	\renewcommand{\algorithmicrequire}{\textbf{Input:}}
	\renewcommand{\algorithmicensure}{\textbf{Output:}}
	\begin{algorithmic}[1]
		\REQUIRE The number of UEs $K$ in the $j$ subarray, the number of subarray antennas $N_{ts}$, the inverse of the SNR $\xi \geq 0$, the subarray channel matrix $\textbf{H}_j^j$, and the number of algorithm iterations $T$;              
		\ENSURE $\textbf G_j$ from \eqref{eq:16}.
            \STATE \textbf{Initiation:} $\textbf A_j \gets 0$;
            \STATE Get $\textbf H_{r,t},$ from \eqref{eq:7}, $\textbf H_{r,t}$ from \eqref{eq:14} and \eqref{eq:15}.
		  \FOR{$idex1=1$ to $K_j$}

               \FOR{$idex2 = 1$ to $N_{ts}$}
             
               \STATE $\textbf H = \textbf H_{r,t}(:,:,idx1,idx2)$;
                    \IF{$\textbf H == 0$}
                   \STATE break;
                  \ENDIF
               \STATE Define $m_j^t\in {\mathbb{C}^{N_{ts}}}$, $n_j^t\in {\mathbb{C}^{K_j}}$ with $m_j^t = 0$, $n_j^t=0$.
               \STATE Define user canonical basis $e_k \in {\mathbb{R} }^{K_j}$, where $[e_k]_k = 1 $and $[e_k]_j = 0, \forall j \neq k$;
                     \FOR{$idex3 = 1$ to $T$}
                     \STATE Select the $r_t$-th row of $(\textbf H_j^j)^H $with  $r_t\in \{1, 2, \dots, K_j\}$ satisfy $P_{r_t}^j = \frac{1}{K_j}$

               \STATE $\eta^t= \frac{[e_k]_{r_t-<h_{jr_t}^j,m_j^t>}- n_{jr_t}^j}{||h_{jr_t}^j||_2^2·}$;
                 \STATE Update $m_j^{t+1}=m_j^{t}+\eta^t h_{jr_t}^j$ ;  
                 \STATE Update $n_{jr_t}^{t+1}=n_{jr_t}^{t}+\eta^t$ ;
                 \STATE Update $\textbf A_j$ from \eqref{eq:20}
                    \ENDFOR
               \ENDFOR

            \ENDFOR
             \RETURN  $\textbf G_j$
	\end{algorithmic} 
\end{algorithm}

Considering that the data symbols of each user are i.i.d. and follow a Gaussian distributed, the instantaneous uplink  $SINR_k^s$ of user $k$ regarding subarray $s$ can be defined as \eqref{eq:23}.
\subsection{Complexity Analysis}

According to the proposed SwoR-rKA algorithm in \cite{b12} shows higher performance in the system and likewise faces higher complexity. In this paper, in order to maintain a high level of performance and low complexity, we choose rKA as precoding against MUI.

 In this subsection, we analyze and compare the complexity of  ZF, rKA and SwoR-rKA respectively, by means of floating point operations(FLOPS) computation, \ $\textbf H\textbf H^H, \textbf H\in {\mathbb{C}^{{n}\times m}}$ requires $n^2m$ FLOPS. 
Assume that $N_t=N=n,S=4$. Combining to classic ZF needs $2n^2m$ FLOPS and rK needs $Im$ FLOPS, $m$ represents the number of iterations of rKA.

\begin{table}[h]
\begin{center}  
   \caption{Computational Complexity for Precoding Methods based on Complex Operations.}  
    \renewcommand{\arraystretch}{2.5}
    \begin{tabular}{c|c|c}  
        \hline
        \hline  
        \textbf{Algorithm} & \textbf{FLOPS} & \textbf{XL-MIMO-AFDM} \\   
        \hline  
        ZF & $N^2*S*2K_s^2N_{ts}$ & $65536N^2S $\\  
        rKA & $N^2S*N_{ts}*T_s$ & $12800N^2S $ \\  
        SwoR-rKA & \makecell{$N^2S \times (N_{ts} \times T_s$ \\ $+ 2N_{ts} \times K_s$)} & \makecell{$16896N^2S$} \\  
        \hline  
    \end{tabular}
   
    \label{tab:1}  
\end{center}
\end{table}
According to Table \ref{tab:1} we can know that rKA has low complexity under XL-MIMO-AFDM system, then the performance of rKA will be presented in the next section.

\section{SIMULATION RESULTS}
In this paper, we focus on the application of our proposed efficient algorithm in XL-MIMO-AFDM systems, comparing the optimized total rate and communication performance with BER.

\begin{figure}[t]
\centering
\includegraphics[width=0.5\textwidth]{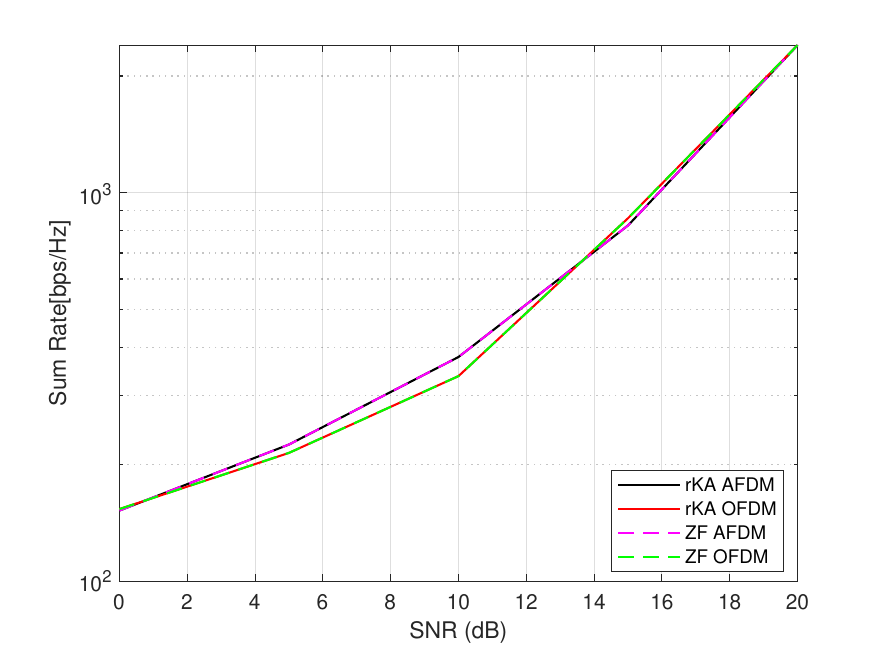}
\caption{Sum-rate performance of a XL-MIMO-AFDM system
employing different precoding schemes$(N_t = 256,K = 32,N_k = 1, S =
4)$.}
\label{fig:3}
\end{figure}

In the first example, shown in \figureref{fig:3}, we can see that our proposed algorithm has lower complexity and better performance compared to \cite{b13}. Consider a BS configured with $N_t = 256$ transmitting antennas broadcasting data to $K = 32$ users with $S = 4$ subarrays and each user configured with $N_k = 1 $ receiving antenna. Simulations show that our proposed rKA has lower complexity and better performance compared to the ZF mentioned in \cite{b13}.

It can be observed that the performance of AFDM is not much improved compared to the sum rate performance of OFDM, and thus verified on simulation in \cite{b13}. Also our proposed rKA algorithm has a better performance than \cite{b13}. However, it is worth noting that while the sum rate provides theoretically achievable bounds, more practical metrics should be considered for a fair comparison between XL-MIMO AFDM and OFDM.

\begin{figure}[t]
\centering
\includegraphics[width=0.5\textwidth]{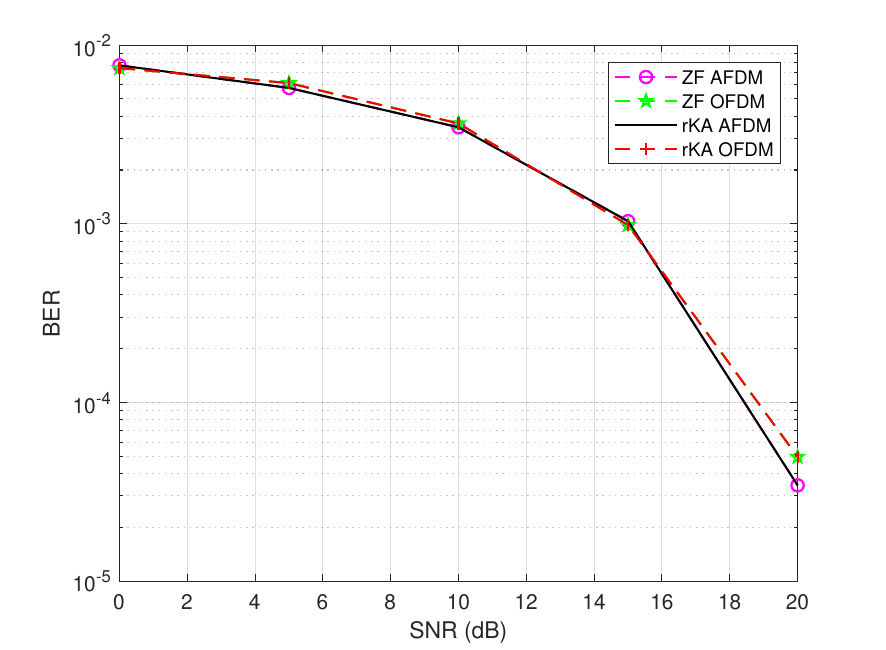}
\caption{BER versus SNR (dB) for spatially precoded AFDM compared with OFDM$(N_t = 256,K = 32,N_k = 1, S = 4)$.}
\label{fig:4}
\end{figure}

In the second example, shown in \figureref{fig:4}, we evaluate the BER of the proposed algorithm. Consider a base station configured with $N_t = 256$ transmitting antennas broadcasting data to $K = 32$ users with $S = 4$ subarrays and each user configured with $N_k = 1 $receiving antenna. Compared to ZF \cite{b13}, our proposed rKA has a lower BER. rKA-AFDM and ZF-AFDM exhibit similar BER, and rKA-OFDM and ZF-OFDM exhibit same BER.

Simulation results show that rKA has lower complexity and superior performance compared to the ZF of \cite{b13}. Also compared to \cite{b5}, \cite{b11}, and \cite{b12}, AFDM can show better performance compared to OFDM in motion scenarios.
Meanwhile, the BER based on ZF-AFDM and rKA-AFDM in the simulation is lower, and the performance is optimized by about 1 dB.

\section{CONCLUSION}
In this paper, we primarily focus on the precoding design under the XL-MIMO-AFDM system model. The application of XL-MIMO significantly enhances system capacity, leading to substantial performance gains for the system in the motion scenarios we aim to design for. Due to the challenge of Doppler frequency offset in dynamic environments, we have implemented AFDM as a solution to mitigate this issue. To address the problems of Multi-User Interference (MUI) and the complex precoding design in the XL-MIMO-AFDM system, we employ the rKA precoding algorithm, which strikes a balance by maintaining high system performance while reducing computational complexity.

\vspace{12pt}
\color{red}


\begin{thebibliography}{00}
\bibitem{b1}M. Cui and L. Dai, "Near-Field Channel Estimation for Extremely Large-scale MIMO with Hybrid Precoding," 2021 IEEE Global Communications Conference (GLOBECOM), Madrid, Spain, 2021, pp.
\bibitem{b2}X. Cao, M. Mohammadi, H. Q. Ngo and M. Matthaiou, "RIS-Assisted XL-MIMO for Coexistence of Near-Field and Far-Field Communications," 2024 IEEE Wireless Communications and Networking Conference (WCNC), Dubai, United Arab Emirates, 2024, pp. 1-6.

\bibitem{b3} J. Luo, J. Fan, K. Xie and X. Shi, "Efficient Hybrid Near- and Far-Field Beam Training for XL-MIMO Communications," in IEEE Transactions on Vehicular Technology, vol. 73, no. 12, pp. 19785-19790, Dec. 2024.

\bibitem{b4} J. Zhu et al., "Decentralization of Tomlinson-Harashima Precoding for MU-MIMO System," 2024 IEEE 100th Vehicular Technology Conference (VTC2024-Fall), Washington, DC, USA, 2024.
\bibitem{b5}Liu, Qiuhao et al. “Performance Analysis of XL-MIMO-OFDM Systems for High-Speed Train Communications.” 2023 IEEE International Conference on Communications Workshops (ICC Workshops) (2023): 1741–1746.
\bibitem{b6}W. Huang, L. Xu, H. Zhang, C. Kai, C. Li and Y. Huang, "Structured OFDM Modulation for XL-MIMO System with Dual-Wideband Effects," in IEEE Transactions on Wireless Communications.
\bibitem{b7}A. Bemani, N. Ksairi and M. Kountouris, "AFDM: A Full Diversity Next Generation Waveform for High Mobility Communications," 2021 IEEE International Conference on Communications Workshops (ICC Workshops), Montreal, QC, Canada, 2021, pp. 1-6.
\bibitem{b8}J. Du, Y. Tang, H. Yin, J. Zhu and Y. Zhou, "A Simplified Affine Frequency Division Multiplexing System for High Mobility Communications," 2024 IEEE Wireless Communications and Networking Conference (WCNC), Dubai, United Arab Emirates, 2024, pp. 1-5.


\bibitem{b9}H. Yin, X. Wei, Y. Tang and K. Yang, "Diagonally Reconstructed Channel Estimation for MIMO-AFDM With Inter-Doppler Interference in Doubly Selective Channels," in IEEE Transactions on Wireless Communications, vol. 23, no. 10, pp. 14066-14079, Oct. 2024, doi: 10.1109/TWC.2024.3408458.

\bibitem{b10}K. R. R. Ranasinghe, H. S. Rou, G. T. F. De Abreu, T. Takahashi and K. Ito, "Joint Channel, Data and Radar Parameter Estimation for AFDM Systems in Doubly-Dispersive Channels," in IEEE Transactions on Wireless Communications.


\bibitem{b11} V. C. Rodrigues, A. Amiri, T. Abrão, E. de Carvalho and P. Popovski, "Low-Complexity Distributed XL-MIMO for Multiuser Detection," 2020 IEEE International Conference on Communications Workshops (ICC Workshops), Dublin, Ireland, 2020, pp. 1-6.

\bibitem{b12}B. Xu, Z. Wang, H. Xiao, J. Zhang, B. Ai and D. W. Kwan Ng, "Low-Complexity Precoding for Extremely Large-Scale MIMO Over Non-Stationary Channels," ICC 2023 - IEEE International Conference on Communications, Rome, Italy, 2023, pp. 6516-6522.
\bibitem{b13}V. Savaux and X. Chen, "Spatial Precoding in Frequency Domain for Multi-User MIMO Affine Frequency Division Multiplexing," 2024 32nd European Signal Processing Conference (EUSIPCO), Lyon, France, 2024, pp. 2112-2116.

\bibitem{b14}V. Savaux, "Special Cases of DFT-Based Modulation and Demodulation for Affine Frequency Division Multiplexing," in IEEE Transactions on Communications, vol. 72, no. 12, pp. 7627-7638, Dec. 2024.









\end{thebibliography}
\end{document}